\documentclass[useAMS,usenatbib]{mn2ermc}
\usepackage{graphicx}
\title[Self-Consistent Analysis of OH Zeeman Observations]{Self-Consistent Analysis of OH Zeeman Observations}

\author[R.~M. Crutcher, N. Hakobian and T.~H. Troland]{Richard M. Crutcher,$^{1}$\ Nicholas Hakobian$^{1}$ and Thomas H. Troland$^{2}$\\
$^{1}$Astronomy Department, University of Illinois, Urbana, IL 61801 USA\\
$^{2}$Physics and Astronomy Department, University of Kentucky, Lexington, KY 40506 USA}

\begin{document}

\date{Accepted 2009 December 9. Received 2009 December 7; in original form 2009 October 13}

\pagerange{\pageref{firstpage}--\pageref{lastpage}} \pubyear{2009}

\maketitle

\label{firstpage}

\begin{abstract}

\cite{CHT} used OH Zeeman observations of four nearby molecular dark clouds to show that the ratio of mass to magnetic flux was smaller in the $\sim0.1$ pc cores than in the $\sim1$ pc envelopes, in contradiction to the prediction of ambipolar diffusion driven core formation. A crucial assumption was that the magnetic field direction is nearly the same in the envelope and core regions of each cloud. \cite{MT} have argued that the data are not consistent with this assumption, and presented a new analysis that changes the conclusions of the study. Here we show that the data are in fact consistent with the nearly uniform field direction assumption; hence, the original study is internally self-consistent and the conclusions are valid under the assumptions that were made. We also show that the Mouschovias \& Tassis model of magnetic fields in cloud envelopes is inconsistent with their own analysis of the data. However, the data do not rule out a more complex field configuration that future observations may discern.

\end{abstract}


\section{Introduction}

The relative importance of magnetic fields and turbulence in driving the formation of molecular cores is a central question in current star formation theory. Measuring magnetic field strengths in order to infer ratios of mass to magnetic flux ($M/\Phi$) has been a focus of observational efforts to answer this question. However, the only technique for directly measuring interstellar magnetic field strengths, the Zeeman effect, has only yielded results for the line-of-sight component $B_{LOS}$ of the magnetic vector {\bf B}. A statistical analysis carried out with several assumptions has been the standard analysis technique for Zeeman results, but this only determines the mean or median value of  \textbar{\bf B}\textbar$\:$for the observed set of clouds, which significantly limits tests of the theory. 

To overcome this limitation, \cite{CHT} (hereinafter CHT) carried out OH Zeeman observations toward the envelope regions surrounding four molecular dark cloud cores, selected from a survey of 34 cores for having strong $B_{LOS}$, and evaluated the {\em ratio} ${\cal R}$ of $M/\Phi$ in a cloud core to that in the envelope. The goal was to test published strong magnetic field models that have uniform initial fields. The value of this technique is that published models of core formation driven by ambipolar diffusion have strong, regular magnetic field morphology such that the unknown angle $\theta$ between {\bf B} and the line of sight is approximately the same in core and envelope regions, allowing  $\theta$ to be eliminated from the expression for ${\cal R}$. The CHT analysis depended on the assumptions that the magnetic field direction was essentially uniform in each cloud and that the ratio of strengths of the quasi-thermal OH lines between envelope and core gave an accurate indicator of the ratio of column densities between envelopes and cores. The idealized ambipolar diffusion theory of core formation requires ${\cal R}$ to be approximately equal to the inverse of the original subcritical $M/\Phi$, or ${\cal R} > 1$. CHT were able therefore to directly test this prediction. They found that the probability that all four of the clouds have ${\cal R} > 1$ is $3 \times 10^{-7}$; the results are therefore significantly in contradiction with the hypothesis that all four of these cores were formed by ambipolar diffusion. 

\cite{MT} (hereinafter MT) have strongly criticized the CHT result, arguing both that the CHT analysis is internally inconsistent and that a different analysis technique that they apply to the CHT data shows that ${\cal R} > 1$ is consistent with the data. In this letter we show that the CHT analysis is internally consistent, and that the MT analysis is itself internally inconsistent.

\begin{table*}
 \centering  
 \begin{minipage}{140mm}
 \caption{Observational Results.}
  \begin{tabular}{@{}lccccccc@{}}
  \hline
   Position  &   $B_{LOS}$ $(\mu$G)  & $(n-s)/\sigma$ & $(n-e)/\sigma$ & $(n-w)/\sigma$ & $(s-e)/\sigma$ & $(s-w)/\sigma$ & $(e-w)/\sigma$ \\
 \hline
L1448n & $-9\pm13$ & 1.5 & 0.1 & 0.3 & 2.4 & 0.8 & 0.4 \\
L1448s & $+14\pm8$ \\
L1448e & $-11\pm 6$ \\
L1448w & $-7\pm7$ \\
\\			
B217n & $-13\pm9$ & 1.4 & 1.7 & 1.7 & 0.3 & 0.2 & 0.1 \\
B217s & $+9\pm13$ \\
B217e & $+5\pm6$ \\
B217w & $+6\pm8$ \\
\\			
L1544n & $-3\pm4$ & 0.5 & 0.4 & 3.6 & 0.3 & 1.7 & 3.2 \\
L1544s & $+2\pm10$ \\
L1544e & $-1\pm4$ \\
L1544w & $+22\pm6$ \\
\\			
B1n & $-16\pm6$ & 0.7 & 1.6 & 1.4 & 1.2 & 0.9 & 0.4 \\
B1s & $-10\pm5$ \\
B1e & $+0\pm7$  \\
B1w & $-3\pm6$ \\
\hline
\end{tabular}
\end{minipage}
\end{table*}

\section{Consistency of the CHT Analysis}

CHT measured $B_{LOS}$ at the four cardinal envelope positions (labelled n, s, e, w) surrounding each core, but they inferred the mean $B_{LOS}$ in each envelope by fitting simultaneously over all the Stokes I and V spectra from the envelope. This enabled them to synthesize a toroidal telescope beam that sampled the envelope while excluding the core. $B_{LOS}$ for the core was measured separately by fitting spectra from a single beam that covered the core area. They then calculated ${\cal R}$ from
\begin{equation}
{\cal R} \equiv \frac{M_{core}/\Phi_{core}}{M_{envelope}/\Phi_{envelope}}.  \label{eq: R}
\end{equation}
The mass (the OH lines are optically thin) and magnetic flux are given by 
\begin{equation}
M \propto I \; \Delta V  \quad  {\rm and}  \quad \Phi \propto (B_{LOS}/cos~\theta),  
 \end{equation}
where $I$ is the line intensity, $\Delta V$ the line width, and $\theta$ the angle between {\bf B} and the line of sight.

With the assumption, based on published strong {\bf B} models (see CHT for references), that $\theta_{core} \approx \theta_{envelope}$, the $cos~\theta$ terms in the numerator and denominator of ${\cal R}$ cancel, and one is left with directly observable quantities only. Thus ${\cal R}$ and its uncertainty could be evaluated. CHT evaluated the uncertainty with a Monte Carlo analysis that utilized the known uncertainties in $I$, $\Delta V$, and $B_{LOS}$. This analysis hinged on one crucial assumption -- that the mean magnetic field direction in the envelope immediately surrounding a core putatively formed by ambipolar diffusion had the same (or nearly the same) direction as in the envelope. MT argued that the data are demonstrably inconsistent with this assumption and that therefore the CHT analysis is invalid. 

\begin{figure}
\includegraphics[width=84mm]{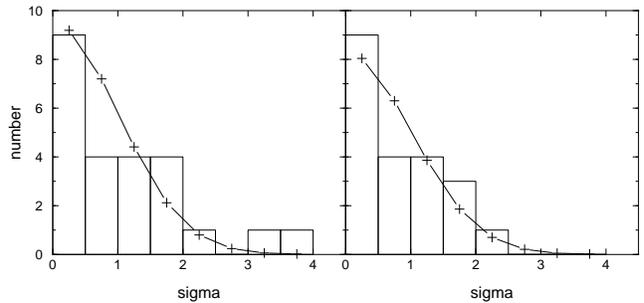}
\caption{Left: Histogram plot shows the difference data from Table 1, (position 1 - position 2)/$\sigma$, in $0.5\sigma$ bins; plus signs show the normal error function for 24 data points. Right: Same as left, except the three differences produced by the L1544w position have been removed, and the normal error function is for 21 data points.}
\end{figure}

Table 1 and Figure 1 show the relevant data. MT claim that the $B_{LOS}$ shown in Table 1 show clear variation in field strength and perhaps even reversals in field direction among the four envelope positions for each cloud. They then argue that this variation violates the CHT assumption that the field direction is essentially uniform in direction for each cloud. 

We also give in Table 1, after the north envelope $B_{LOS}$ entry for each cloud, each of the six possible differences between envelope values of $B_{LOS}$, divided by the $1\sigma$ uncertainty in each difference from the individual measurement uncertainties. If MT are correct, these data should show a scatter significantly greater than that imposed by the measurement uncertainties. However, the mean of the 24 differences is $1.12\pm0.25$, where the uncertainty takes into account that there are 16 and not 24 independent measurements. The mean for the normal error function is 0.80. Hence, the scatter in the $B_{LOS}$, including the nominal reversals in direction, is only marginally larger than the scatter that would result from measurement uncertainty. Most of the larger scatter is produced by the single position L1544w; excluding this position, the mean of the remaining 21 differences is $0.87\pm0.18$, in agreement with measurement error being entirely responsible for the scatter in the 15 measured $B_{LOS}$ excluding the L1544w position. The differences are also shown graphically in Figure 1, along with the normal error function for comparison. If the scatter in the $B_{LOS}$ were due entirely to measurement uncertainty, these two plots would agree. These plots confirm the conclusion drawn from the mean of the differences. Even with the L1544w position included, the measured differences agree fairly well with the normal error function, while if L1544w is excluded, the agreement is excellent. Hence, while the L1544w position may be anomalous -- as discussed originally by CHT -- there is no statistically significant evidence for the MT claim that the CHT assumption of a fairly uniform field direction for each cloud is invalid. The scatter in the differences is entirely attributable to the measurement uncertainties and not to any intrinsic scatter in the $B_{LOS}$. Even if the L1544 cloud were excluded from the CHT analysis, the CHT conclusion that these cores were not formed by ambipolar diffusion remains valid. 

MT give an example of possible measurements of 10 $\mu$G and 14 $\mu$G, each with uncertainty 0.1 $\mu$G, and note that the mean differs from each value by 2 $\mu$G, not the 0.07 $\mu$G given by propagation of errors. However, these $100\sigma$ and $140\sigma$ examples are not germane to the CHT case of roughly $1-2\sigma$ measurements. Moreover, CHT did not average the four envelope results for each cloud and obtain the uncertainty by error propagation; they synthesized a toroidal beam to sample the envelopes and obtained the uncertainties directly from the single envelope $B_{LOS}$ measurement for each cloud. 

\section{Consistency of the MT Analysis}

MT argue that an arbitrarily twisted field morphology (see the cartoon shown in MT Figure 1) must be included in the analysis for ${\cal R}$. Although we have argued above that the data do not require that such a morphology is present, let us follow MT and assume that it is. The MT analysis of the CHT data should then be consistent with this proposed model of the field morphology -- that is, that the angle $\theta$ between the core and envelope fields are arbitrarily large. But then the MT analysis {\em itself} is internally inconsistent with this model. CHT defined ${\cal R}$ in order to eliminate the unknown angle $\theta$ between the field direction and the line of sight; CHT assumed that between the core and the envelope of a cloud those directions are the same (except for minor differences in the $\theta$ that do not significantly affect the analysis, see discussion in CHT). That assumption allowed the unknown angle $\theta$ between the field and the line of sight, which enters as $cos~\theta$, to drop out of the ratio ${\cal R}$ (see Equations 1 and 2). If the $\theta$s for the four envelope positions and the core of each cloud vary greatly, as suggested in MT Figure 1, then $\theta$s do not drop out of ${\cal R}$. For a self-consistent MT analysis, each of the five different $\theta$s (core and four envelope) would have to be explicitly included in the expression for ${\cal R}$. However, MT do not do so; such an expression would have the five unknown $\theta$s and could not be evaluated. Instead, MT use our expression for ${\cal R}$ with the $cos~\theta$s missing. MT stated that they were only allowing for different {\em magnitudes} in {\bf B} over the four envelope positions, not for different directions $\theta$. But this assumption is completely inconsistent with the astrophysical motivation of strongly twisted field lines (MT Figure 1 and discussion) that they give for rejecting the CHT analysis and substituting their own. MT offer no astrophysical explanation for fields at the envelope positions varying significantly in strength and perhaps even being antiparallel while the angle $\theta$ remains invariant. The MT analysis is therefore not self-consistent, and cannot be used to analyze the CHT data. 

Even within the framework of the MT analysis, it appears that the uncertainty in the ${\cal R}s$ is overestimated. MT considered the variation of the measured $B_{LOS}$ as one component of the uncertainty, and then added as a second component the measurement uncertainties. However, as we showed above, the variation in the measured $B_{LOS}$ is consistent with being due to the measurement uncertainties and not to a real variation. MT appear to be doubly counting the uncertainties.

\section{Conclusion}

CHT concluded that their measurements of the ratios of $M/\Phi$ between envelopes and cores did not agree with the prediction of the ambipolar diffusion model. Here we have shown that the CHT analysis is internally self consistent; their conclusions are valid within the framework of the assumptions they made. The validity of the MT paper rests on two pillars: (1) that the CHT data analysis procedure is unambiguously inconsistent with the data itself, and (2) that MT have a superior analysis technique. We have demonstrated that neither of these pillars of their paper is correct. The conclusions of \cite{CHT} therefore stand -- the observed variations of $M/\Phi$ from envelope to core are not consistent with the prediction of the ambipolar diffusion driven theory of star formation. This conclusion does not, of course, rule out the possibility that there are structures in magnetic field morphology near dark cloud cores; higher resolution and higher sensitivity observations would be necessary to investigate this possibility. 

The approach of CHT to test the ambipolar diffusion driven model of star formation by measuring the change in $M/\Phi$ between envelope and core is a powerful one that should be further exploited, since it reduces uncertainties in actual values of magnetic field direction and mass estimates by taking ratios. Unfortunately, such experiments will require very large amounts of telescope time. However, use of the eVLA for OH Zeeman mapping and ALMA for CN Zeeman mapping may make it possible to extend this technique to smaller scales without requiring such large assignments of telescope time.

\section*{Acknowledgments}

This work is partially supported by the NSF under grants AST 0307642, 0606822 and 0908841.

\end{document}